\documentclass[a4paper]{jpconf}
\usepackage{graphicx}

\begin{document}
\title{Weak signal extraction using matrix decomposition, with application to ultra high energy neutrino detection}
\author{S Prohira, for the T576 collaboration}
\address {Center for Cosmology and AstroParticle Physics (CCAPP), The Ohio State University, Columbus OH, USA}

\ead{prohira.1@osu.edu}
\begin{abstract}

In radio-based physics experiments, sensitive analysis techniques are often required to extract signals at or below the level of noise. For a recent experiment at the SLAC National Accelerator Laboratory to test a radar-based detection scheme for high energy neutrino cascades, such a sensitive analysis was employed to dig down into a spurious background and extract a putative signal. In this technique, the backgrounds are decomposed into an orthonormal basis, into which individual data vectors (signal + background) can be expanded. This expansion is a filter that can extract signals with amplitudes $\sim$1~\% of the background. 
This analysis technique is particularly useful for applications when the exact signal characteristics (spectral content, duration) are not known. In this proceeding we briefly present the results of this analysis in the context of test-beam experiment 576 (T576) at SLAC.

\end{abstract}

\section{Introduction}

In this paper we discuss an analysis technique applied to data taken as part of test-beam experiment 576 (T576) at the SLAC National Accelerator Laboratory in 2018. The aim of T576 was to establish a technique for remote detection of ultra high energy (UHE) neutrinos. When an UHE neutrino interacts in a dense material (such as ice) it will produce a cascade of particles, moving relativistically. As these particles move through the material they produce ionization. For high primary energies, this ionization can become dense enough to reflect incident radio~\cite{krijnkaelthomas,radioscatter,krijn_radar_18}. Depending upon the lifetime of the ionization, which is an open experimental question, this radio reflection may be detectable by remote receivers. If so, it would be a viable detection method for UHE neutrinos beyond the reach of current optical detectors~\cite{icecube}. Such a measurement would be the first of its kind. 

For the T576 experiment, a beam of $\sim10^9$ electrons at 10~GeV was directed into a target of high-density polyethylene (HDPE) in the End Station A facility at SLAC. Meanwhile, a radio transmitter was broadcasting continuous wave (CW) radio in the direction of the target. Several receiving antennas were also directed toward the target, and monitored with a digital oscilloscope. The purpose of T576 was to attempt a measurement of the radio reflection from the shower created when the electrons traverse the target. The density of the shower should be roughly equivalent to that of a $10^{19}$eV primary neutrino, making it a good proxy for an in-nature scenario. 

The backgrounds in this experiment, however, far exceeded those that would be experienced in nature. The primary background was transition radiation produced as the charged electrons passed from the vacuum of the beam pipe into air, and then from air into the HDPE target. This would not be expected in nature because a) the primary in nature, a neutrino, is not charged and b) the antennas are embedded in the same material as the shower. This large background was $\cal{O}$(100x) the amplitude of our expected signal, and so an advanced analysis technique was required to dig into this background. We therefore developed a method using matrix decomposition that allowed for very thorough filtration of the data, and removal of the spurious background. It is similar, though distinct, from several existing matrix decomposition analysis techniques~\cite{karhunen, loeve, stat_sig_proc, pca_review, svd_radiometry}, and is based closely upon~\cite{bean_ralston}. We present here the technique being used in the ongoing analysis.

The paper is organized as follows. We will first explain the analysis technique in detail, outlining the mathematical formalism. We will then discuss the cross-checks that were performed to verify the technique, and finally  quantify its sensitivity in this particular application.

\section{Mathematical formalism}

\subsection{Matrix decomposition}

Matrix decomposition is a method used to lower the dimensionality of data, reducing it into a series of {\it modes} or {\it patterns} (both terms will be used in this document, following the formalism of~\cite{bean_ralston} and~\cite{t576_run1}). These patterns are orthogonal, and for what we describe here, form a basis into which other vectors can be expanded, e.g.,

\begin{equation}\label{basis}
  V_j=c^{\alpha}e^{\alpha}_j,
\end{equation}
where some vector $V$ is shown as an expansion of basis modes $e$ with coefficients $c$. The label $\alpha$ indexes the modes, and $j$ indexes the vector elements. Here the $j$th element is shown as a sum over the labels $\alpha$, with summation implied over repeated indices. For orderly data, a small number of modes $e$ contain the majority of the information within a given vector.

Matrix decomposition finds the basis modes $e$. For the present application, a matrix $\mathbf{M}$ is composed of $n$ column vectors $V$, each of length $m$. Each of these vectors is a triggered radio event, a series of sampled voltages. In general, if a matrix $\mathbf{M}$ is composed from $n$ vectors $V$ then any vector $V$ within $\mathbf{M}$ can be described completely by Equation~\ref{basis}, with the coefficient $c$ found via

\begin{equation}
  c^{\alpha}=V_je^{\alpha}_j.
\end{equation}

There are many ways in which the matrix $\mathbf{M}$ can be decomposed, but here we will consider singular value decomposition (SVD), which has the form

\begin{equation}
  \mathbf{M}=\mathbf{u}\mathbf{\Lambda} \mathbf{v}^*,
\end{equation}
where $\mathbf{u}$ is an $m\times m$ matrix, $\mathbf{v}$ is an $n\times n$ matrix, and $\mathbf{\Lambda}$ is a diagonal $m\times n$ matrix with `singular values' down the main diagonal, which are the relative weights with which the corresponding rows and columns of $\mathbf{u}$ and $\mathbf{v}$ contribute to $\mathbf{M}$, and are generally shown in decreasing order.

\subsection{Building a basis}

The decomposed matrices can be used to build a basis of patterns $e$, as above. To get these patterns, we zero all of the singular values and put a $1$ in the position we wish to expand,

\begin{equation}
  \mathbf{\Lambda}^{\alpha}_{ii}=\delta_{i\alpha},
\end{equation}
and then we get the filter matrix for the label $\alpha$ via

\begin{equation}
  \mathbf{M}^\alpha=\mathbf{u}\mathbf{\Lambda}^{\alpha}\mathbf{v}^*
\end{equation}
which we then sum over the $n$ dimension to get the basis pattern $e^{\alpha}$,

\begin{equation}
  e^{\alpha}_j=\sum_{i=1}^n \mathbf{M}^{\alpha}_{ji}
\end{equation}

The more orderly the data used to construct $M$, the fewer patterns $e$ required to reconstruct the individual vectors $V$. In the limit that every column of the matrix $M$ is identical, there will only be one basis mode with non-zero elements, and it will be identical to the columns of M. In the other limit, where the vectors $V$ (the columns of $M$) are pure, uncorrelated noise, then none of the basis modes $e$ will, alone, be able do describe any vector $V$.

\subsection{Making a filter}

The next step in the procedure is to produce a filter to extract signal from high-background data. To do this, two datasets are needed. The first dataset is the `real' dataset, which contains a putative signal, plus background, plus noise. The second is the `null' dataset, which contains only background plus noise. Then the basic procedure is simple:

\begin{itemize}
\item Produce a filter basis, as above, from the null dataset. This set contains background and noise only, but no signal.
\item For each real vector $V^r$, construct a filter $f^n$ by expanding $V^r$ in the filter basis to order $n$.
\item The filtered vector $V^f$ is attained by subtracting the filter from the data vector. 
\end{itemize}

This procedure is best performed on normalized data. The norms can be retained and re-applied after filtration to achieve the correct scaling. The process is conceptually similar to Fourier expansion, but instead of modes of cosines and sines of different frequencies, the modes are detailed patters constructed from the backgrounds present in the null set. If properly acquired, a filter composed from the null basis should allow for almost complete filtration of the real data, such that only signal and noise remain. We will now present an example of the method.

\begin{figure}[ht]
  \begin{centering}
    \includegraphics[width=.5\textwidth]{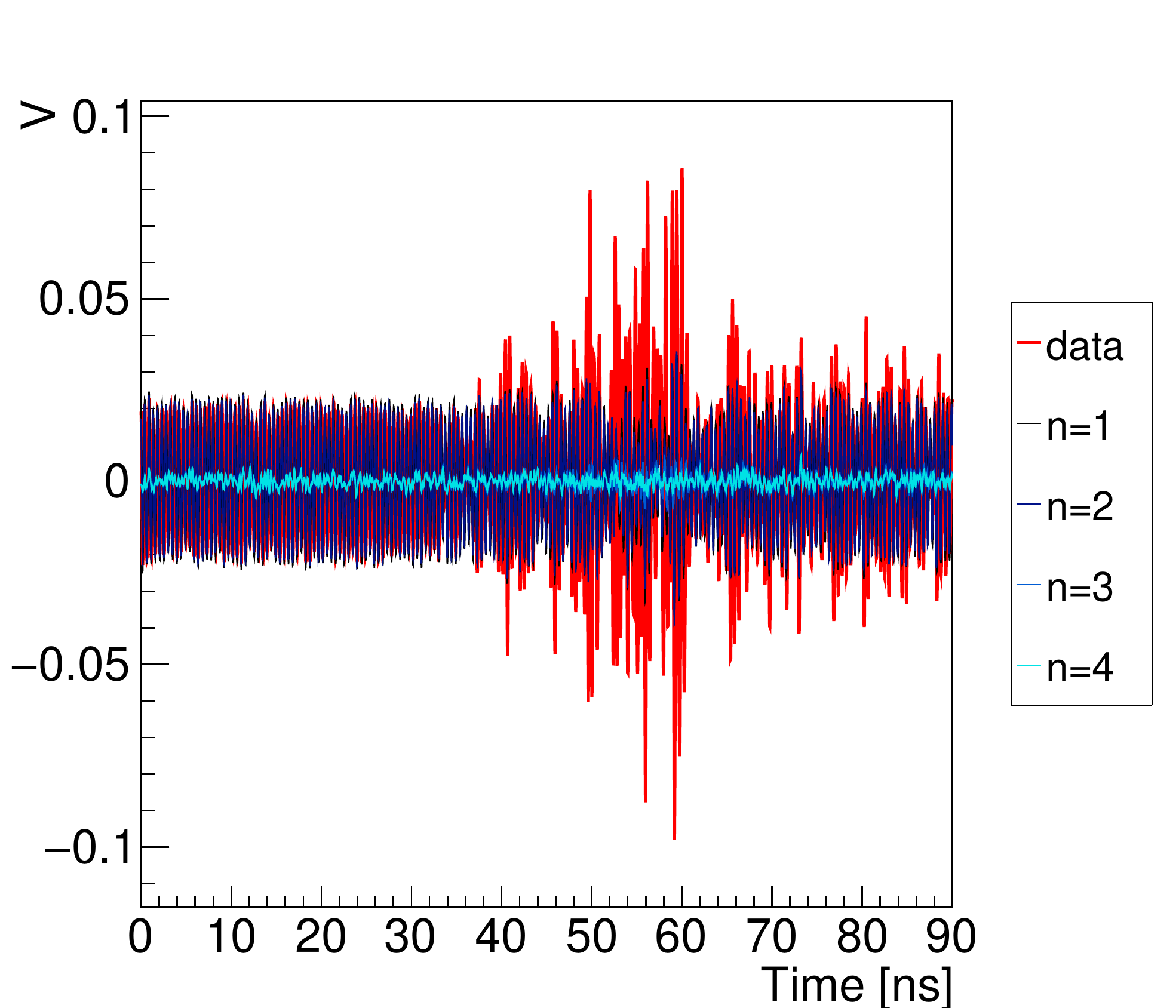}
    \par\end{centering}
  \caption{An example `null' data event (red online), overlaid by the same event after filtration via an SVD filter made from an increasing number of basis modes.}
  \label{fig:expansion}
\end{figure}

\section{Example expansion in a basis}

\subsection{Filtration of null data}

In what follows, we will be using actual background waveforms (`null' data) from the T576 experiment. These waveforms are constructed so as to not include signal. During T576, CW was broadcast constantly toward the target, and when the particle bunch entered the target, this CW was reflected from the particle shower ionization out to the receivers. This reflection (if it is strong enough to be detected) is our signal. The CW received by the receiving antennas, before the shower happens, is therefore free of signal. As the electron bunch exits the beam pipe and enters the target, high-amplitude radio is produced from several mechanisms: sudden appearance~\cite{sa}, transition radiation~\cite{tr, krijn_eas}, and Askaryan radiation~\cite{askaryan_orig}. This is present in each event, and we call it `beam splash'. A radar signal could only be present when the CW was on and beam was present. So, to produce `null' data, we take pre-signal CW and add a beam splash event for which the transmitter was off. Such null data mimics the real data to high precision. In this section we demonstrate the basis expansion method, using this background-only data. In following sections we will inject simulated signals.

\begin{figure}[ht]
  \begin{centering}
    \includegraphics[width=.5\textwidth]{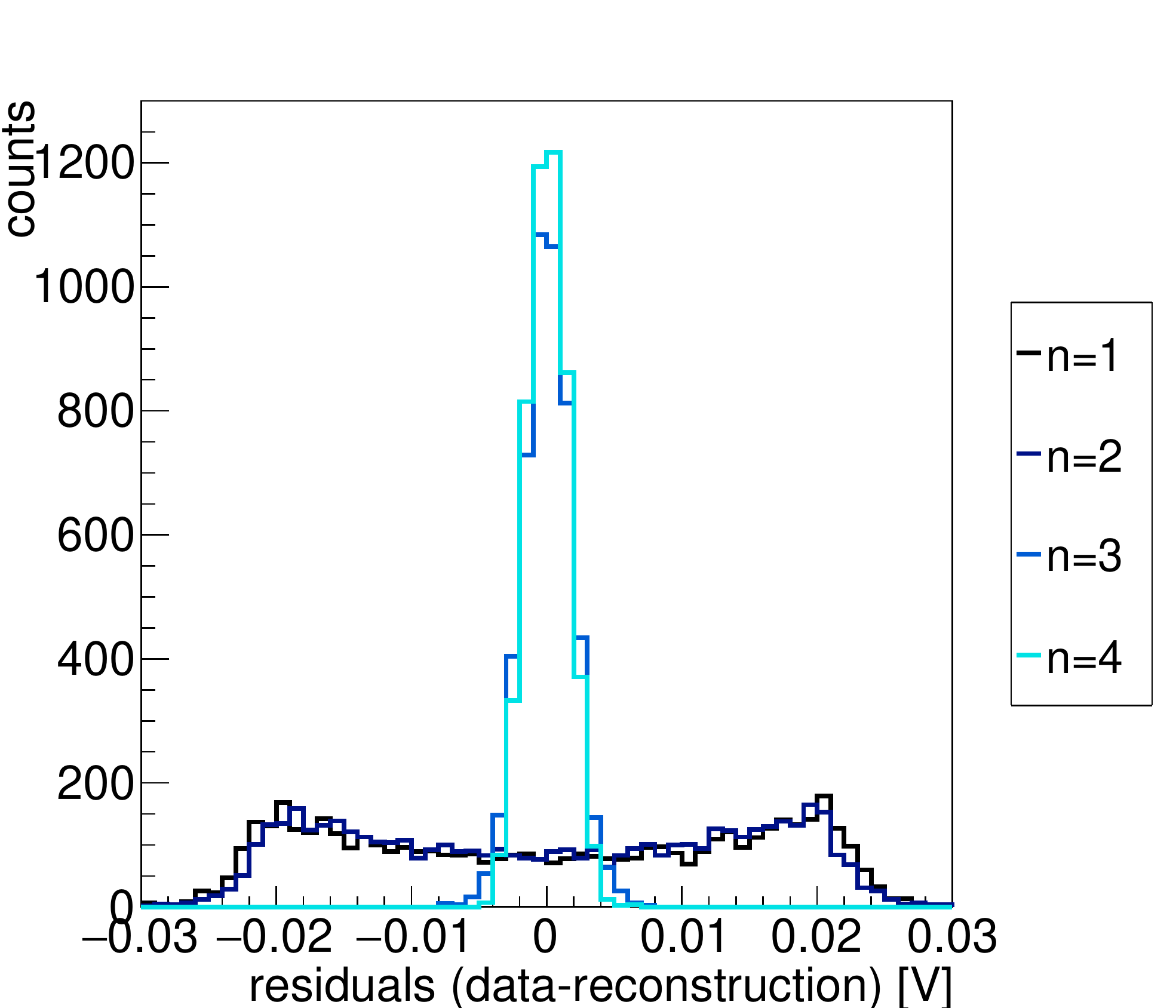}
    \par\end{centering}
  \caption{The residuals of the filtration process using an increasing number of basis modes.}
  \label{fig:residuals}
\end{figure}

Figure~\ref{fig:expansion} shows a typical waveform from the T576 experiment in the background. Using 50 similar waveforms, we constructed a basis via the above procedure. This particular event in Figure~\ref{fig:expansion} is {\it not} one of the events used to construct the basis. Overlaid are the $V^f$ for what's left over after filtration in increasing number of basis modes. That is, the traces labeled $n=$1, 2, 3, and 4 are the filtered waveforms after performing $V^r-f^n$ for $n=$1, 2, 3, and 4. After expansion in only 4 modes, the filtered waveform seems to be purely noise, as it should be, since we have no signal here. Figure~\ref{fig:residuals} shows the residuals $V^r_j-f^n_j$, showing convergence to a narrow distribution around zero, which should be the noise distribution of the SLAC facility where these data were taken, all that is left in $V^f$ after filtration.

\subsection{Extraction of an injected signal}

To test the sensitivity of the method, we can inject a signal into null data and extract it via the above procedure. This gives a measure of how capable this technique is to extract extremely small signals. We use a completely independent set of null events for the following signal-extraction analysis as we did for the construction of the basis, which has been built following the above procedure. That is, we built two independent sets of null data. One set was used to construct the filter basis, and the other set was injected with signal, as below. 

\begin{figure}[h]
  \centering
  \includegraphics[width=.95\textwidth]{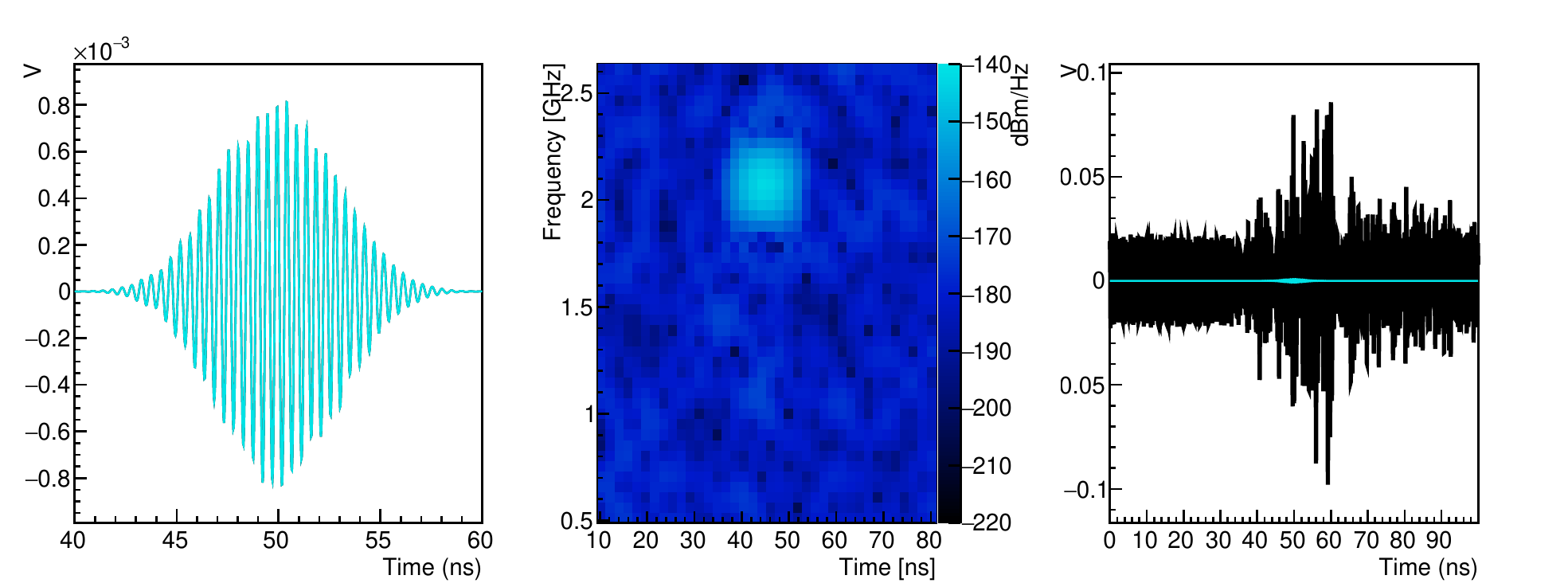}
  \par\caption{A simulated signal injected into the null data, which will be extracted via SVD filtration. Left: the time-domain signal. Center: a spectrogram of the signal, showing the frequency content. Thermal noise has been added for clarity. Right: the signal in context. To make the fake event, these two curves are added.}
\end{figure}\label{fig:simsig}

We make a very simple signal out of random-phase CW, windowed with a Gaussian window, and scaled to a peak voltage of approximately 1\% the peak voltage of the background in Figure~\ref{fig:expansion}. Such a signal is a simplified version of what a radar signal might look like, and is shown in both the time domain and the time/frequency domain 
in Figure~\ref{fig:simsig}. Also shown in this figure is the pulse in context, overlaid on top of the null event to which it will be added. The simulated pulse is added to the null event to make a fake event.

\begin{figure}[h]
  \centering
  \includegraphics[width=.9\textwidth]{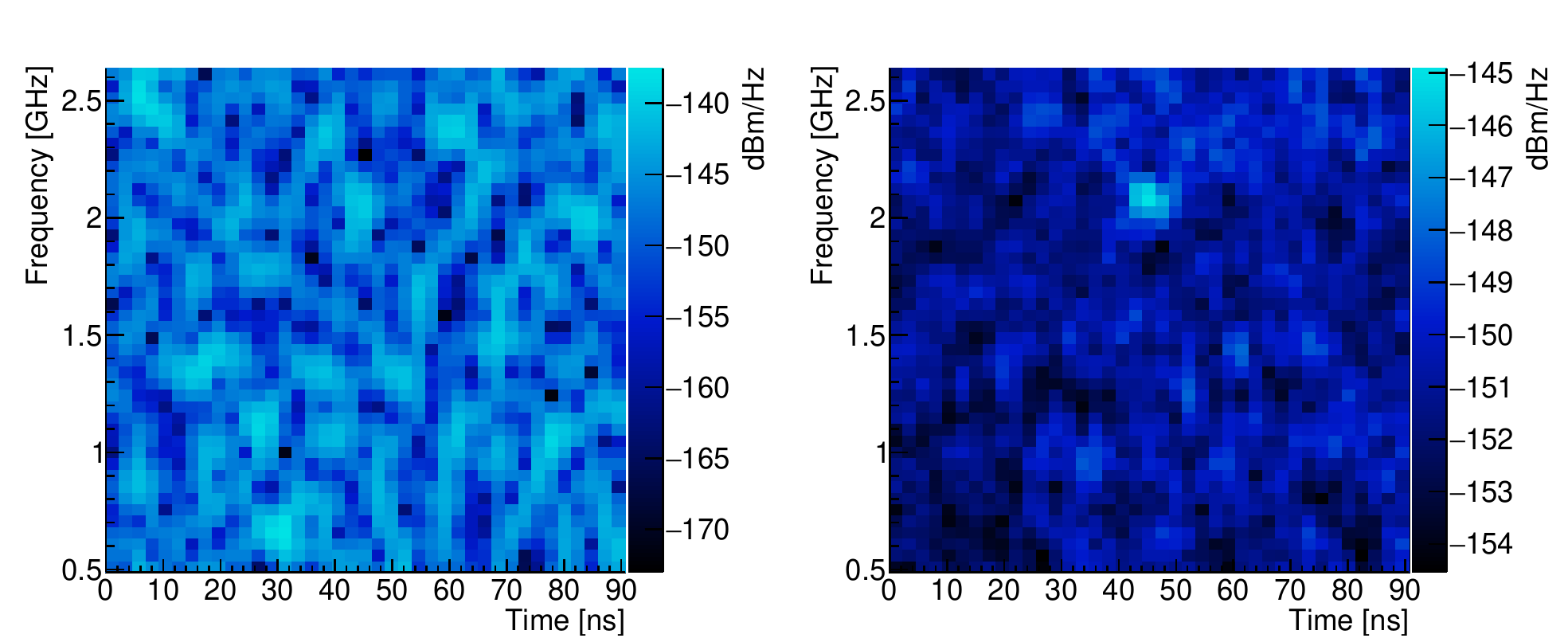}
  \par\caption{Left: a spectrogram of the result of filtering on a single event. Here, signal is difficult to observe by-eye above noise. Right: an average spectrogram of 100 similarly filtered events (each with random phase, injected into different null events, also with random CW phase, and SVD-filtered). There is a clear signal excess in the average spectrogram.}
  \label{fig:result}
\end{figure}

We then build a filter for this event by expanding it in the null basis, and subtracting the filter from the event itself, as above. The left panel of Figure~\ref{fig:result} shows the resultant spectrogram for the filtered event after this procedure. The signal in this event is difficult to discern by eye above the level of noise in this representation, but since now all that remains is signal and noise, any number of techniques could be used to extract the signal, including averaging over similar events. To that end, we repeat this procedure 100 times on different events, injecting signals with the same shape and timing, but random phase, into random phase null events. Each of these events is expanded in the filter basis as above, then filtered, and the resultant spectrograms are averaged. This average of all filtered events is shown in the right side of Figure~\ref{fig:result}, showing a clear signal excess. 
The peak amplitude of this signal is comparable to that of the pure simulated signal in the central panel of Figure~\ref{fig:simsig}, indicating minimal loss of signal through filtration. 

It is therefore evident from this procedure that 1\% signals can be extracted from noise even when both the phase of the background and the phase of the signal are random. Such a method is advantageous in an in-nature radar-based experiment, where the distances to real neutrino induced cascades is not known ahead of time, and therefore the received signal will always have a random phase relative to the sounding CW. 

\section{Application}

The procedure herein has been employed in the analysis of the second run of data taking for T576, which concluded in November of 2018. In the analysis of the first run data, an excess consistent with a radar signal was extracted via a similar technique~\cite{t576_run1}. The present analysis is a more robust method, as shown here, capable of digging deeper into large backgrounds than the analysis from that first run. We therefore present this proceeding as a detailed explanation of the method which we employ in our forthcoming analysis.

\section{Conclusion}

Filtration via matrix decomposition is a powerful technique for extracting a small signal within a large background, even when this background has random phase and fluctuating amplitude. This technique can extract small signals at the 1\% level, and can be used to advantage in radio-based physics applications. 

\section{Acknowledgements}

We express our gratitude to the SLAC National Accelerator Laboratory test beams personnel, for providing us with invaluable on-site support and excellent beam conditions for T576. This research was partially funded by a US Department of Energy Office of Science Graduate Student Research (SCGSR) award. The SCGSR program is administered by the Oak Ridge Institute for Science and Education for the DOE under contract number DE-SC0014664. 

\section*{References}
\bibliographystyle{unsrt}
\bibliography{/home/natas/Documents/physics/tex/bib}

\end{document}